\documentclass[journal]{IEEEtran}
\usepackage[table,xcdraw]{xcolor}
\usepackage{amsfonts}
\usepackage{cite,graphicx,amsmath,amsthm}
\usepackage{amsmath}
\usepackage{subcaption}
\usepackage{caption}
\usepackage{fancyhdr}
\usepackage{dsfont}
\usepackage{array}
\usepackage{bm}
\usepackage{float}
\usepackage{algorithm}
\usepackage{algpseudocode}
\usepackage{multirow}
\usepackage{booktabs}
\usepackage{multirow}
\usepackage{makecell}
\usepackage{tikz}
\usetikzlibrary{positioning} 
\usepackage{hyperref}
\usepackage{graphicx} 
\usepackage{wrapfig}

\usepackage{tikz} 
\usepackage{diagbox}
\usepackage[T1]{fontenc}
\begin{document}

	\title{Large Language Model Agents for Radio Map Generation and Wireless Network Planning}
	
	\author{Hongye Quan, Wanli Ni, \textit{Member, IEEE}, Tong Zhang, \textit{Member, IEEE},  Xiangyu Ye,  Ziyi Xie, \textit{Member, IEEE}, \\  Shuai Wang,    \textit{Senior Member, IEEE},
			Yuanwei Liu, \textit{Fellow, IEEE}, and Hui Song \textit{Member, IEEE}
		
		\thanks{H. Quan, X. Ye, and Z. Xie are with the Harbin Institute of Technology, Shenzhen, 518055, China  (e-mail: \{200210402,210210416\}@stu.hit.edu.cn; xieziyi@hit.edu.cn).}
		\thanks{T. Zhang is with the Harbin Institute of Technology, Shenzhen, 518055, China, and also with the National Mobile Communications Research Laboratory, Southeast University, China (e-mail: tongzhang@hit.edu.cn).}
		\thanks{W. Ni is with the Department of Electronic Engineering, Tsinghua University, Beijing 100084, China (e-mail: niwanli@tsinghua.edu.cn).}
		\thanks{S. Wang is with the Shenzhen Institute of Advanced Technology, Chinese Academy of Sciences, Shenzhen 518055, China (e-mail: s.wang@siat.ac.cn).}
		\thanks{Y. Liu is with the Department of Electrical and Electronic Engineering, The University of Hong Kong, Hong Kong (e-mail: yuanwei@hku.hk).}
		\thanks{H. Song is with the Ranplan Wireless Network Design Ltd., Cambridge, UK (e-mail: hui.song@ranplanwireless.com).}
	}
	
	\maketitle

	\begin{abstract}
		Using commercial software for radio map generation and wireless network planning often require complex manual operations, posing significant challenges in terms of scalability, adaptability, and user-friendliness, due to heavy manual operations. To address these issues, we propose an automated solution that employs large language model (LLM) agents. \textcolor{black}{These agents are designed to autonomously generate radio maps and facilitate wireless network planning for specified areas, thereby minimizing the necessity for extensive manual intervention.} To validate the effectiveness of our proposed solution, we develop a software platform that integrates LLM agents. Experimental results demonstrate that a large amount manual operations can be saved via the proposed LLM agent, and the automated solutions can achieve an enhanced coverage and signal-to-interference-noise ratio (SINR), especially in urban environments. 
	\end{abstract}
	

	\section{Introduction}	
	In the rapidly evolving landscape of telecommunications, the demand for efficient, scalable, and reliable wireless network services become quite strigent \cite{Letaief,Saad}. The deployment and optimization of wireless networks require precise radio map generation to ensure optimal coverage and performance \cite{WeiZhang,Bakirtzis}. Traditional methods, such as interpolation \cite{Dongsoo}, ray-tracing algorithm \cite{Suga}, and artificial intelligence (AI) model \cite{Bakirtzis-TAP},  for creating these maps and planning network infrastructure are often characterized by their complexity, high costs, and large amount of training data. A conventional and state-of-the-art way, adopted by network operators, for implementing radio map and network planning is by using commercial software.  However, when encountering increasingly dense and dynamic urban environments in future 5G-A/6G networks, the use of commercial software may face difficulties in scalability, adaptability, and user-friendliness, due to heavy manual operations in commercial software. 
	
	Most recently, the advent of large language model (LLM) has opened new possibilities across various domains, including natural language processing, medicine, and mathematical discoveries \cite{medicine,Mathematical,Ren}.
	LLM excels at pattern recognition, predictive analytics, and understanding unstructured data,  making them a promising tool for addressing some of the challenges inherent in traditional wireless network planning processes \cite{Qiu,Chatzistefanidis,Sevim,Jingwen}. Qiu \textcolor{black}{\emph{et al.}} \cite{Qiu} optimized wireless network planning with LLM-enabled optimization, where the results showed that LLM can effectively optimize the wireless network planning.	
	By leveraging the advanced capabilities of LLM agents, it becomes possible to develop more automated, intelligent systems that can generate accurate radio maps and plan wireless networks with minimal human oversight. 
	For example, Chatzistefanidis \textcolor{black}{\emph{et al.}} \cite{Chatzistefanidis} proposed a collaborative framework that leveraged LLMs to facilitate a higher level of abstraction and automation in network planning, which could significantly streamline the deployment process.
	Sevim \textcolor{black}{\emph{et al.}} \cite{Sevim} took a different approach by designing a reinforcement learning-based system augmented with LLM capabilities for wireless network deployment.
	This approach underscored the potential of LLMs in dynamically adjusting network parameters to meet varying operational requirements.
	Additionally, Tong \textcolor{black}{\emph{et al.}} \cite{Jingwen} explored the application of LLM agents in network slicing, revealing the models' ability to handle complex tasks within wireless networks.
	Despite these advancements, the specific application of LLM agents for radio map generation and network planning remains an underexplored area.

	\begin{figure*}[h]
		\centering
		\includegraphics[width=6.4in]{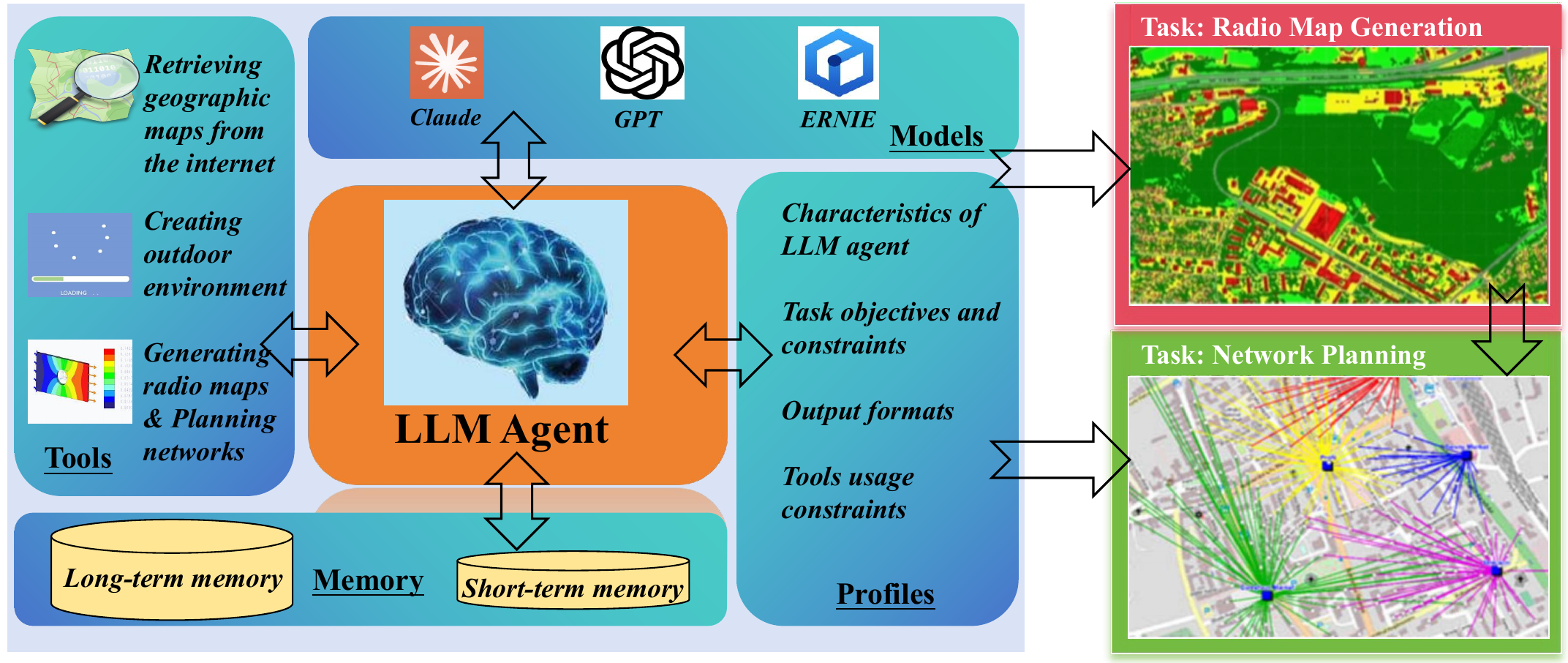}
		\caption{Proposed framework of LLM agent for radio map generation and wireless network planning.} \label{F1}
	\end{figure*}

	In this letter, for the first time, we investigate and demonstrate how LLM agents improve radio map generation and wireless network planning. 
	The main contributions of this work are summarized as follows.
	\begin{itemize}
		\item 
		\textcolor{black}{We propose an innovative framework that integrates LLM agents to automate the generation of radio maps and the deployment strategy of base stations (BSs).}
		Our framework is designed to scale effectively across different geographic regions and environmental conditions, providing robust support for both existing and emerging network technologies.
		\item 
		We develop a software platform that utilizes LLM agent to drive the commercial software, so as to predict signal propagation patterns and select optimal locations for BSs.
		Experimental results on real-world scenarios, show that a large amount of manual operation can be reduced by LLM agents, compared to the conventional manual operations without LLM agents.
	\end{itemize}


	\section{The Proposed LLM Agent Framework}
	
	An LLM agent framework is proposed for automating radio map generation and wireless network planning tasks, as shown in Fig. \ref{F1}. The proposed framework has three modules: Tools, Models, and Profiles. The \textbf{Tools} module plays a critical role, as it distinguishes the LLM agent from a standard LLM by enabling the use of external tools. In this framework, we develop four specific tools: 1) retrieving geographic maps from the internet, 2) creating outdoor environment models, 3) generating radio maps through dedicated scripts, and 4) planning wireless networks via specialized scripts. The \textbf{Models} module serves as the application programming interface (API) for the LLM, facilitating remote access to advanced models such as GPT-4, Claude, ERNIE, and others. Finally, the \textbf{Profiles} module is designed to define the specific mission and characteristics of the LLM agent, tailoring its behavior to the task at hand.

	Once the above three modules are equipped, the proposed LLM agent is able to automatically execute radio map generation and wireless network planning tasks according to the input prompts. Specifically, this LLM agent has two functionalities, i.e., task planning, and memory. Task planning functionality refers to executing  radio map generation and wireless network planning tasks with reasoning and planning.  The memory functionality refers to enhancing the task planning by efficiently managing a list of executed and pending tasks for preventing duplication.  Below, we provide more details and explanations on the three modules and two functionalities.

	\subsection{The Module Design}
	
	\subsubsection{Profile Module}

	The profile of this proposed LLM agent is built to ensure efficient task planning and execution,  
	and user interaction. The LLM model is designed for general purpose. Hence, there is a need to define the 
	specific mission and characteristics of proposed LLM agent through a profile file. 
	
	In particular, the profile of our proposed LLM agent should require the character of LLM agent, task objective and constraints, output format, file path, and tool usage constraints. 
	For the character of LLM agent, we can appoint LLM agent as both radio map generator and wireless network planner. Moreover, the task objective and constraints, output format, file path, and tool usage constraints are regulated in details, so that LLM agent can understand the objective,  restrictions, and requirements on its tasks. For example, the character of LLM agent can be defined as "You are a powerful radio map generation and wireless network planning assistant capable of performing simple operations in the \textit{name of wireless network planning software} using available tools." The task objective and constraints, output format, file path, and tool usage constraints can be defined according to radio map generation and wireless network planning tasks.

	\subsubsection{Tool Module}
	
	The key feature of LLM agent is the ability of using tools. As such, tools should be  carefully and efficiently designed.
	For completing radio map generation and wireless network planning tasks, related tools are elaborated as follows. We plan to use special software for radio map generation and wireless network planning. To package specified functions of software into tools, we can write scripts to control the software functions. In particular, there are three tools. A tool getting geographic map from internet, such as downloading from OpenStreetMap (OSM) website\footnote{\url{https://www.openstreetmap.org/}}, should be constructed. A tool importing the downloaded geographic map to special software for outdoor environment creation is needed. A tool performing radio map generation or wireless network planning is necessary. The LLM agent then knows how to use the above three tools with planning and memory, which will be explained later on.
	
	\subsubsection{Model Module}
	
	To empower the LLM agent,  the LLM model API should be embedded as a module. The prompts are putted into LLM model through API. Then, LLM model conduct reasoning and  returns a log file containing parameters and usage for tools, on the condition that the profile is known in advance. The regular expression searches the log file to retrieve the parameters and usage for tools from it. By this way, the LLM plays as the head of the agent and provides the intelligence in task planning, and memory, which will be explained later on.
	
		\begin{figure}[t]
		\centering
		\includegraphics[width=0.45\textwidth]{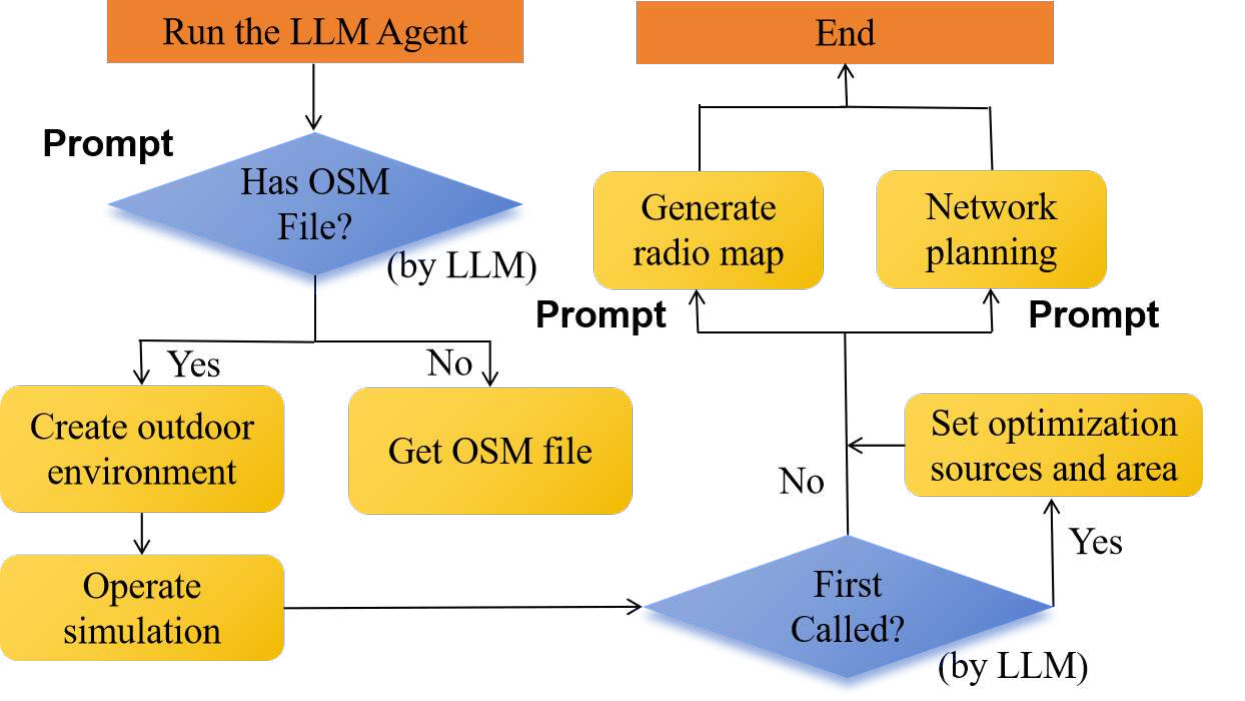}
		\captionsetup{labelfont={color=black}}
		\caption{\textcolor{black}{Flow chart of the proposed software platform, where OSM file refers to OpenStreetMap file.}} \label{FC}
	\end{figure} 
	
	\subsection{Functionality}
	
	\subsubsection{Memory Functionality}
	
	The memory functionality empowers the proposed LLM agent by memorizing past and current data. The memory can be divided into short-term and long-term memory. The short-term memory refers to temporary prompts and  inference results that the LLM agent uses immediately within the current task. The short-term memory is wiped out once the task is complete.\textcolor{black}{The long-term memory helps the LLM agent understand the current prompt by reviewing historical multi-round prompts. The long-term memory refers to prompts and inference results accumulated across multiple tasks, thus avoiding repetition and providing better user experience.} Both shot-term and long-term memory efficiently manages historical and temporary tasks for their prompts and inference results for preventing duplication.

	\subsubsection{Task Planning Functionality}

	Task planning in an LLM agent involves breaking down a complex task into smaller, executable sub-tasks, prioritizing them for efficient execution. 
	The radio map generation task and wireless network planning task are correlated, i.e., the wireless network planning task is based on the result of radio map generation. Therefore, when executing wireless network planning task, the radio map generation task should be executed in priority. According to the input prompts, the LLM agent dynamically adjusts its plan during execution to handle changes and ensures the task is completed. This process is iterative with feedback, to refine and optimize future plans.
	
	\section{Software Platform and Experiment Results}
	
	In this section, we first introduce our developed software platform, and then show and explain our experiment results.

	\begin{figure}[t]
		\centering
		\includegraphics[width=0.45\textwidth]{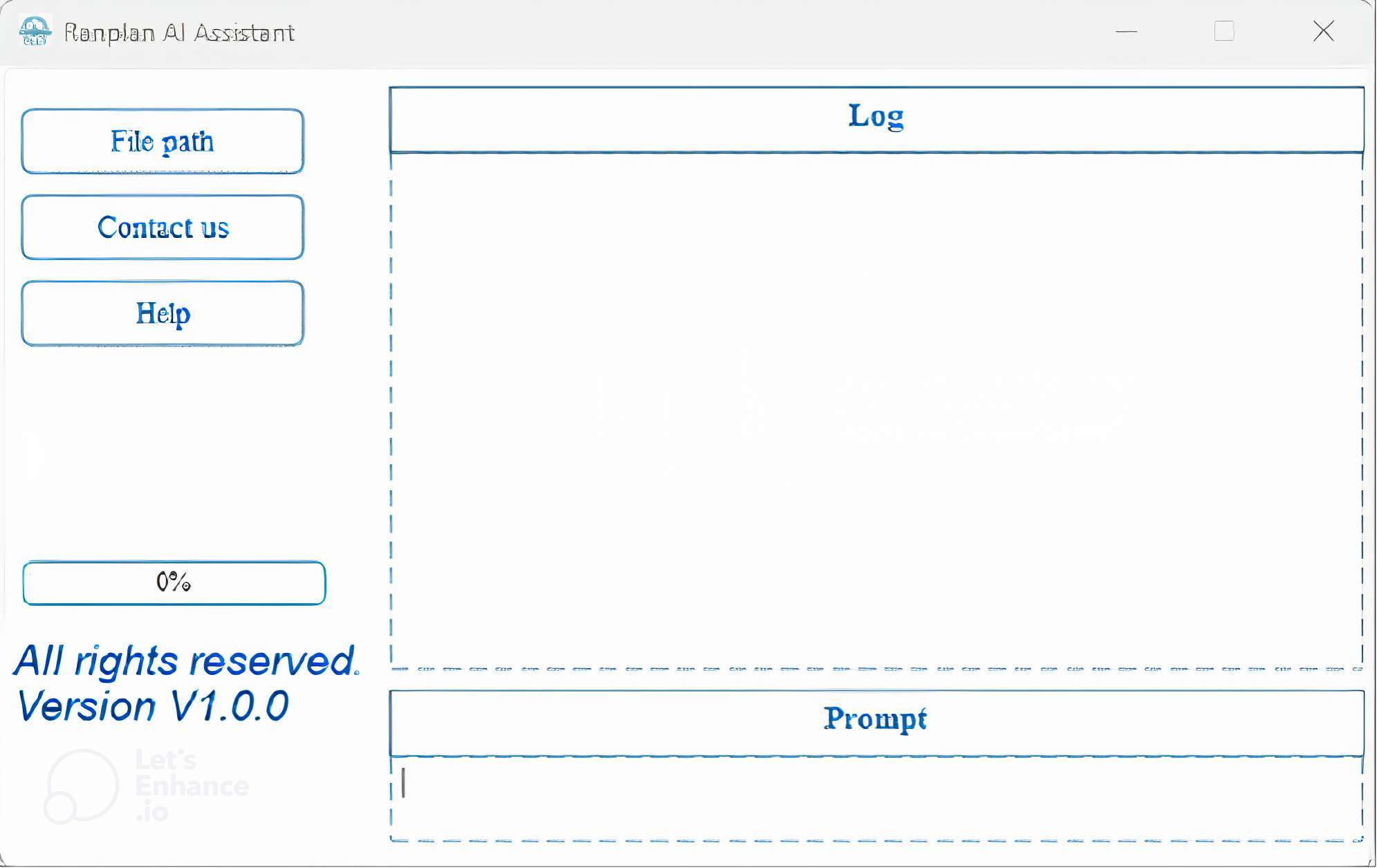}
		\captionsetup{labelfont={color=black}}
		\caption{\textcolor{black}{User interface of the proposed software platform.}} \label{UI}
	\end{figure} 
	
	\begin{figure*}[htbp]
		\centering
		\centering
		\includegraphics[width=0.93\textwidth]{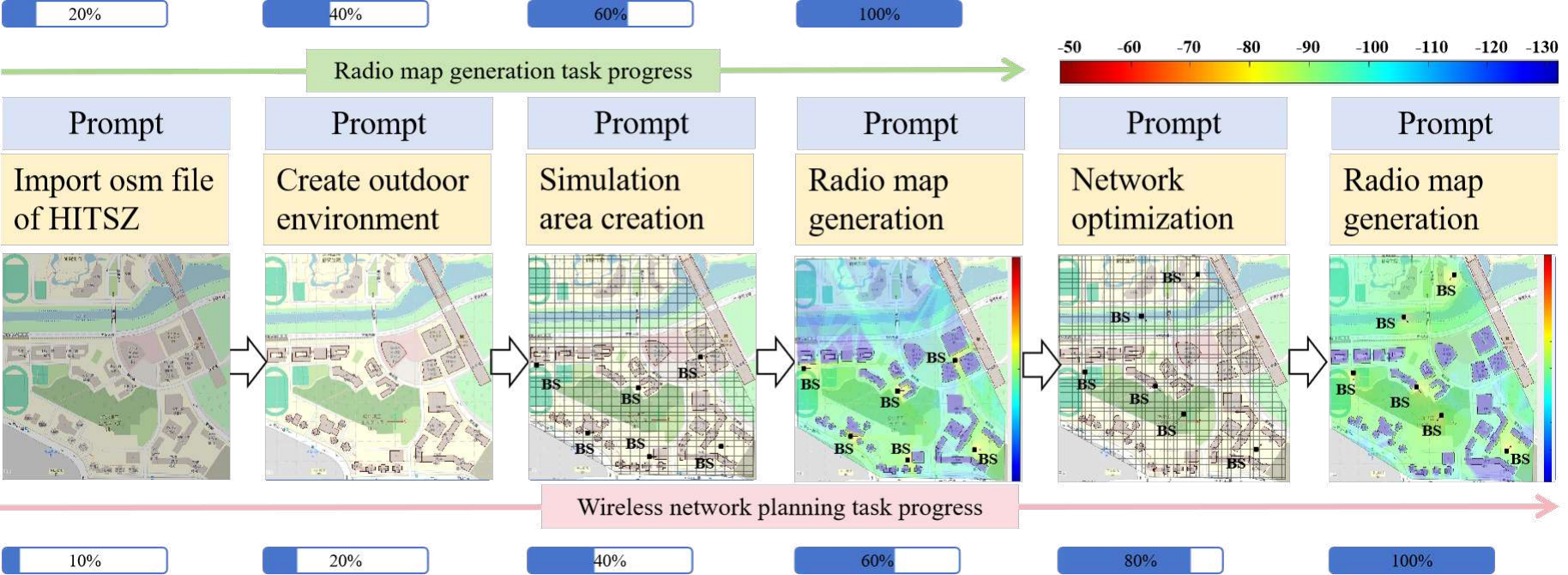}
		\captionsetup{labelfont={color=black}}
		\caption{\textcolor{black}{Radio map generation and wireless network optimization tasks for HITSZ.}} \vspace{0.25cm}
		\label{hitsz}
	\end{figure*}
	
	\begin{figure*}[htbp]
		\centering
		\includegraphics[width=0.93\textwidth]{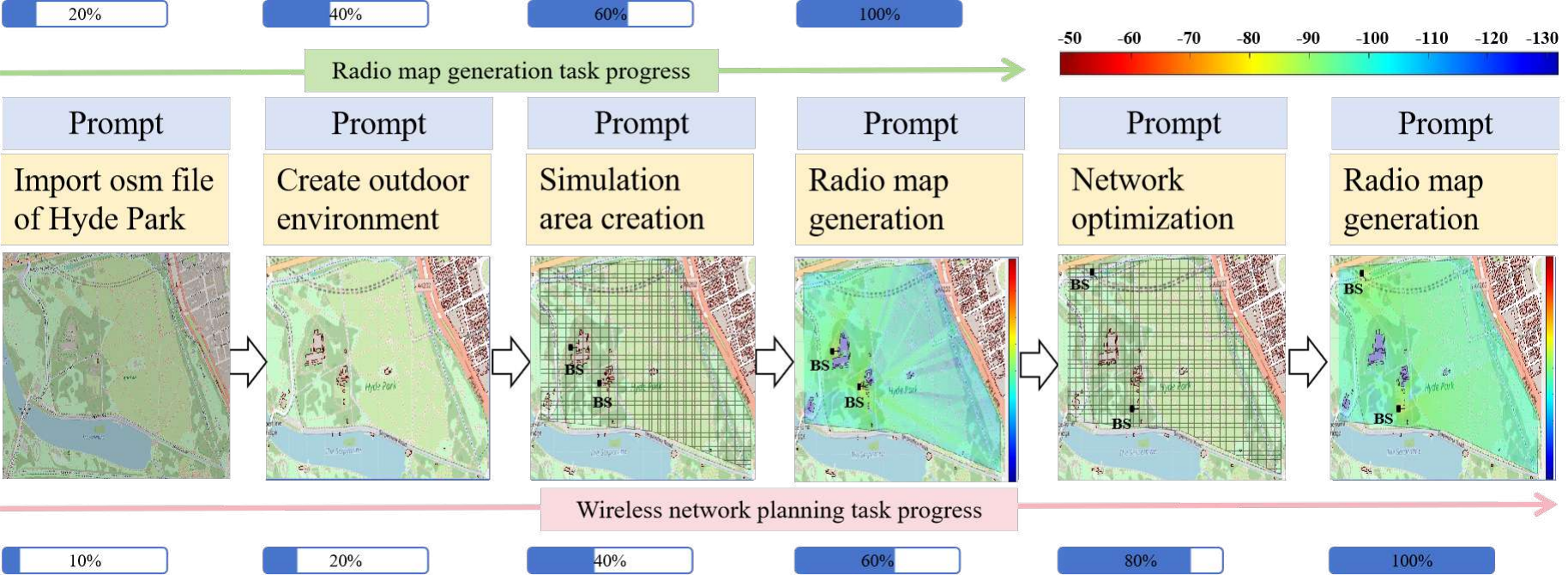}
		\captionsetup{labelfont={color=black}}
		\caption{\textcolor{black}{Radio map generation and wireless network optimization tasks for Hyde Park.}}
		\label{hdp}
	\end{figure*}

	\subsection{Software Platform Development}
	
	To implement the proposed LLM agent, we design a software platform using \textsc{PySide6} and \textsc{Python}. Moreover, the platform has been exported as an executable (.exe) program, simplifying the launch and usage process. The flow chart of the proposed software platform is given in Fig. \ref{FC}. It can be observed from Fig. \ref{FC} that the OSM file is downloaded first, then the environment is created, and finally specific functions related to tasks are executed. All actions are carried out according to the prompts highlighted in Fig. \ref{FC}. 
		
	The user interface (UI) of the designed software platform is shown in Fig. \ref{UI}. The proposed UI comprises the following five functional areas:  1) settings area providing with "File path",   "Contact Us",   and "Help",  2) prompt and log area,  3)  log area,  4) the execution area,  and 5) progress bar area. This software platform serves as a proof-of-concept and provides a basis for more sophisticated designs in the future.

	\subsection{Experiment Results}
	
	To evaluate the performance of the proposed software platform with LLM agent, we perform a series experiments based on real scenario settings.
	The proposed software platform is designed upon RANPLAN ACADEMIC V6.8.0 software\footnote{\url{https://www.ranplanwireless.com/}}, where the later is specifically developed to radio map generation and wireless network planning tasks in academic purpose. The experiment settings are list as follows. As for project configuration, we deploy 5G NR BS operated at 5 GHz with a bandwidth of 80 MHz. The height of placed transmitter is set to the default value of $2.4$ m, with a drive test resolution of $5$ m to capture the spatial differences in outdoor environment. The potential influence of the indoor sources on outdoor network environment is not considered for simplicity.
	
	Figs. \ref{hitsz} and \ref{hdp} illustrate the running processes of the proposed software platform in Hyde Park, England and Harbin Institute of Technology, Shenzhen (HITSZ), China, encompassing the entire process from initial osm file downloading to radio map generation, and final network planning optimization. To be specific, first with the prompt \texttt{"Import osm file of HITSZ/Hyde Park"}, the agent turns to OpenStreetMap website for downloading the geographic maps. 
	Then, with the prompt \texttt{"Create outdoor environment"}, the agent establishes a foundational outdoor scene comprising buildings, green areas, and roads from the downloaded geographic map, providing the geographical basis for subsequent tasks. Following this, the prompt \texttt{"Simulation area creation"} defines a grid-based simulation region where user randomly deploys BSs, enabling precise segmentation of the radio propagation environment. Subsequently, the \texttt{"Radio Map Generation"} operation generates an initial distribution map of radio signal coverage, where varying colors indicate spatial differences in best path-loss (PL) among all BSs, which is a pivotal metric that quantifies the reduction in signal strength. An increase in PL typically corresponds to higher signal attenuation or loss. Automatic cell optimization (ACO) function in RANPLAN ACADEMIC V6.8.0 is an effective solution to this wireless network planning problems. Regarding the wireless network planning, ACO optimizes source placement, count, transmission power, azimuth, and down-tilt angles to enhance signal strength and eliminate redundancies. Compliance criteria of our network planning is set such that the PL does not exceed $100$ dB for $80$\% of the area. To regulate the density of BSs, a minimum spacing of $50$m is enforced between any two BSs. By leveraging the \texttt{"Network Optimization"} prompt, the agent optimizes BS configurations and signal coverage, significantly improving the uniformity and efficiency of the radio coverage area while adjusting the directional orientation of BSs, by using ACO. Finally, another \texttt{"Radio Map Generation"} prompt generates the optimized PL heatmap distribution, validating the effective results of planning.
	
	The observations and insights obtained from Figs. \ref{hitsz} and \ref{hdp} are summarized as follows. The initial BS placement in Fig. \ref{hitsz} reveals that most areas, aside from open areas near the transmitter, are dominated by high PL (indicated in purple), primarily due to the substantial obstruction caused by numerous buildings. On the contrary, the initial BS placement result in Fig. \ref{hdp} shows that few buildings makes the PL less deterioration. Figs. \ref{hitsz} and \ref{hdp} also indicate that the network planning process not only determined the sites for all BSs, but also adjusted the orientation angles of BSs. The ACO function is successful attained its objective as we can observe the changing of radio map color.  To further verify the effectiveness of the LLM agent, we additionally select signal-to-interference-plus-noise ratio (SINR) as a metric for simulation comparison, which will be demonstrated followed. Note that for SINR metric, the best PL times transmit power is the desired and the other PL times transmit power is the interference.
	
\begin{figure}[t]
		\centering
				\begin{subfigure}[t]{.44\textwidth}
			\centering
			\includegraphics[width=\textwidth]{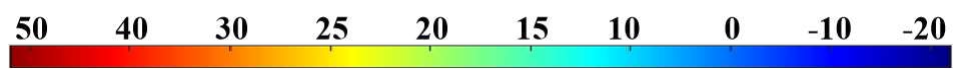}
			\vspace{0.01cm}
		\end{subfigure}
		\begin{subfigure}[t]{0.235\textwidth}
			\centering
			\includegraphics[width=.8\textwidth]{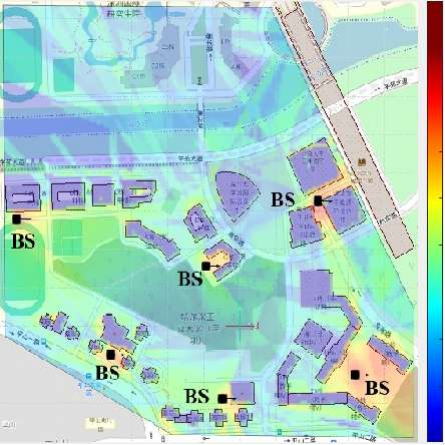}
			\caption{HITSZ before planning.}
			\label{hitsz3}
		\end{subfigure}
		\hspace{0.0cm}
		\begin{subfigure}[t]{0.235\textwidth}
			\centering
			\includegraphics[width=.8\textwidth]{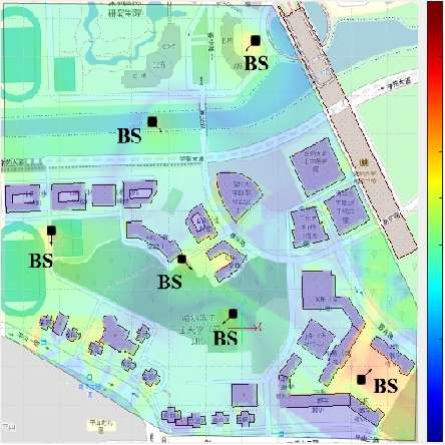}
			\caption{HITSZ after planning.} 	\vspace{0.25cm}
			\label{hitsz5}
		\end{subfigure}	
		\begin{subfigure}[t]{0.235\textwidth}
			\centering
			\includegraphics[width=.8\textwidth]{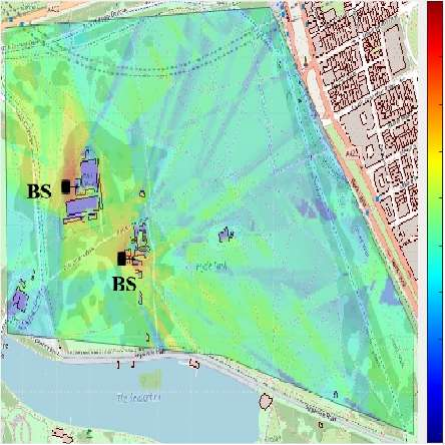}
			\caption{Hyde Park before planning.}
			\label{hdp3}
		\end{subfigure}
		\begin{subfigure}[t]{0.235\textwidth}
			\centering
			\includegraphics[width=.8\textwidth]{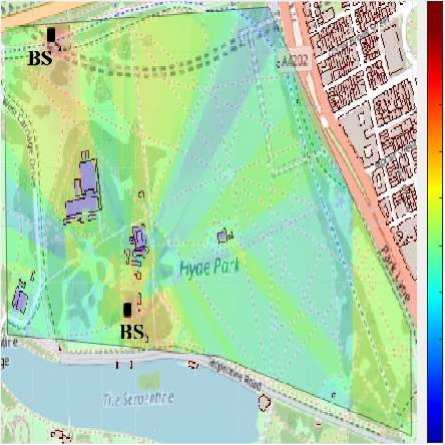}
			\caption{Hyde Park after planning.}
			\label{hdp5}
		\end{subfigure}	
		\captionsetup{labelfont={color=black}}	
		\caption{\textcolor{black}{SINR heatmap (dB) for HITSZ and Hyde Park.}}
		\label{sinr}
	\end{figure}

\begin{figure}[t]
		\begin{subfigure}[t]{.241\textwidth}
			\centering
			\includegraphics[width=\textwidth]{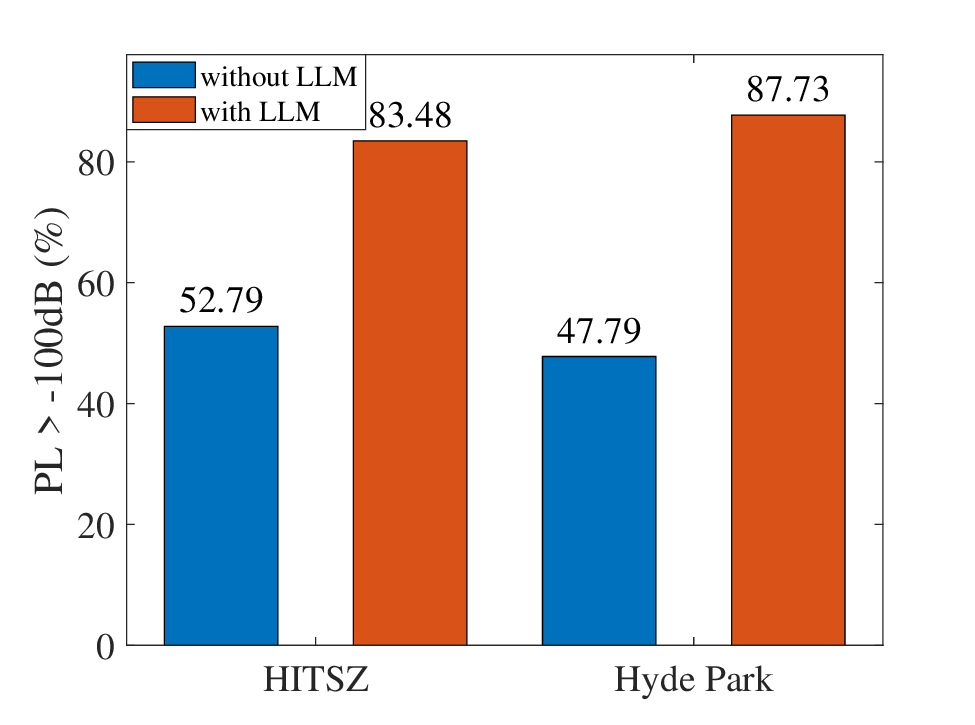}
			\caption{Ratio of PL $>$ $-100$dB area.}
			\label{pl_com}
		\end{subfigure}
		\begin{subfigure}[t]{.241\textwidth}
			\centering
			\includegraphics[width=\textwidth]{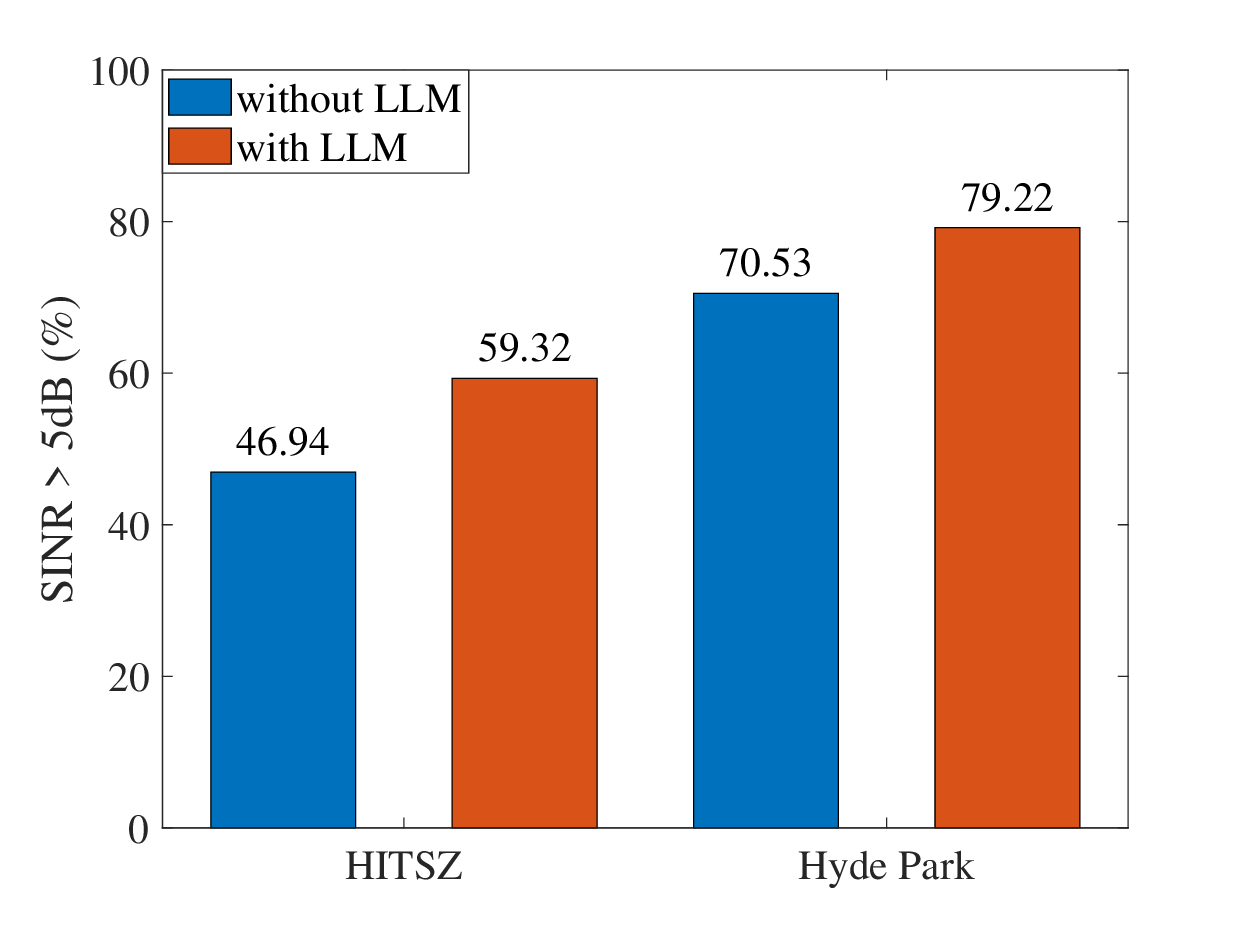}
			\caption{Ratio of SINR\ $\textgreater$ 5dB area.}
			\label{sinr_com}
		\end{subfigure}
		\captionsetup{labelfont={color=black}}
		\caption{\textcolor{black}{Performance evaluation of network planning results.}}
		\vspace{-0.1cm}
		\label{data}
	\end{figure}

	Fig. \ref{sinr} shows SINR heatmaps for two distinct scenarios before and after network planning, illustrating the efficacy of the LLM agent from SINR view of point. Specifically, the transmission power of each BS is set at 1W. Fig. \ref{data} illustrates a more intuitive comparison through statistical data from network planning results in Figs. \ref{hitsz}-\ref{sinr}. For both PL $>$ $-100$dB and SINR $\textgreater$ $5$dB criteria, the coverage percentage with LLM surpassed that without LLM in two locations. Furthermore,	the HITSZ scenario enjoying more SINR gain, since ACO is	more effective in urban environment with numerous buildings.

	Table \ref{tab} shows the user-friendliness and reduction of manual operations by utilizing the software platform with LLM agent. In case of manual operation, the implementation of each function requires individual operations as 6, 9, 20 and 22, \textcolor{black}{where the details of these operations can be found in the user's manual in RANPLAN ACADEMIC V6.8.0} In contrast, with our LLM agent, the whole process can be achieved by entering the prompt only 4 times. That is to say, the LLM agent reduces $91.4$\% and $93.0$\% manual operations for radio map generation and network planning tasks, respectively.  Hence, with the advent of LLM agents, it is possible to automate the radio map generation and wireless network planning in any locations on the earth.

	\section{Conclusions}
	This letter presented the first LLM agent framework and corresponding software platform for radio map generation and wireless network planning tasks.  By means of automating the RANPLAN ACADEMIC V6.8.0 software, the proposed LLM agent significantly enhances task execution efficiency while ensuring high performance in planned wireless networks. \textcolor{black}{In the future, this ground-breaking LLM agent software platform can be utilized for fully automated generation of large-scale radio maps to support AI training, and error feedback will be added.} Moreover, it offers significant value to operators, enterprises, telecom vendors, and system integrators involved in designing, and deploying public and private networks.
	
				\begin{table}[t]
		\centering
		\captionsetup{labelfont={color=black}}
		\caption{\textcolor{black}{Comparison of number of manual operations.}}
		\begin{tabular}{|c|c|c|c|c|}
			\hline
			\diagbox[width=2cm]
			& \makecell[c]{Import \\ osm file} 
			& \makecell[c]{Create \\ environment} 
			& \makecell[c]{Generate \\ radio map} 
			& \makecell[c]{Network \\ planning} \\ \hline
			\makecell{Without any \\ LLM agents} & 6 & 9 & 20 & 22 \\  \hline
			\makecell{With an \\ LLM agent}  &  \cellcolor{gray!20} 1  &  \cellcolor{gray!20} 1  &  \cellcolor{gray!20} 1 &  \cellcolor{gray!20} 1 \\ \hline
		\end{tabular}
		\label{tab}
	\end{table}
	
	\bibliographystyle{IEEEtran}
	\bibliography{NL2024Ref}

\end{document}